# Global Tweet Mentions of COVID-19

Guangqing Chi, Junjun Yin, M. Luke Smith, and Yosef Bodovski

The Pennsylvania State University
Email: gchi@psu.edu

**Abstract**
*Background.* After a year and half and over 4 million deaths, the COVID-19 pandemic continues to be widespread, and its related topics continue to dominate the global media. Although COVID-19 diagnoses have been well monitored, neither the impacts of the disease on human behavior and social dynamics nor the effectiveness of policy interventions aimed at its containment are fully understood. Monitoring the spatial and temporal patterns of behavior, social dynamics and policy—and then their interrelations—can provide critical information for preparatory action and effective response. *Methods.* Here we present an open-source dataset of 1.92 million keyword-selected Twitter posts, updated weekly from January 2020 to present, along with a dynamic dashboard showing totals at national and subnational administrative divisions. *Results.* The dashboard presents 100% of the geotagged tweets that contain keywords or hashtags related COVID-19. We validated our inclusion criteria using a machine learning-based text classifier and found that 88% of the selected tweets were correctly labeled as related to COVID-19. With this information we tested the correlation between tweets and covid diagnosis from January 1, 2020 through December 31, 2020 and see a decreasing correlation across time. *Conclusions.* With emerging COVID variants but ongoing vaccine hesitancy and resistance, this dataset could be used by researchers to study numerous aspects of COVID-19 and provide valuable insights for preparing future pandemics.

## 1. Introduction

The rapidly spreading Sars-CoV-2 (COVID-19) virus has reached every corner of the world [1]. A major challenge in studying the social and behavioral aspects of COVID-19 is that traditional social data collection methods are slow, labor intensive, and expensive. Further, such methods may do a poor job of measuring key population characteristics and behaviors, even when robust probability sampling is used. In contrast, social media data, including those from social networking services (e.g., Twitter, Facebook, and Instagram) provide massive amounts of rich textual data in real time. Of these sources, Twitter provides a highly accessible Big Data stream and has drawn interest from researchers in many disciplines [2]. By studying tweet mentions over space and time, researchers can track phenomena such as public awareness of COVID-19, social distancing, anti-lockdown advocacy, attitudes toward reopening communities, concerns about the economy, anti-Asian sentiment, mental health problems, and others [3,4]. Also, because the dataset is large, investigators can draw meaningful information from small geographic units (e.g., counties, neighborhoods) and fine time intervals (e.g., specific days).

To enable the use of Twitter data for studying the social and behavioral aspects of COVID-19, we developed a dashboard to track global tweets about the disease across time and space [5]. The data come from the 100% geotagged tweets that we have collected continuously since January 2020. A geotagged tweet is denoted with the latitude and longitude of the location from where it was sent. The data include all geotagged tweets with keywords or hashtags related to COVID-19 in English, Spanish, French, Italian, German, Chinese, Korean, Japanese, and other languages. A Twitter user who is concerned about being infected might tweet across multiple days about symptoms [6,7], seeing a health care provider, having a blood test, and/or receiving lab results. It should be noted that Twitter allows developers to access tweets mentioning COVID-19 in real time [8], which provides the most comprehensive collection of COVID-19 related tweets, but this Twitter data stream was not available until May 2020. Further, the data do not provide geographic locations, thereby losing the capacity of linking the data to local contexts. Some researchers have also collected tweets related to COVID-19 and shared the tweet IDs via GitHub repositories. For example, the Panacea Lab at Georgia State University [9] and the Information Sciences Institute of the University of Southern California [10] have been collecting such data, but their repositories contain only a small fraction of the information embedded in each tweet (i.e., tweet ID, post time, and language).

Our team has been continuously collecting geotagged tweets with global coverage of conversations related to COVID-19 since January 2020. Our database also includes other information from each tweet, such as the tweet text and followers. As of July 31, 2021, we have collected 1,921,468 geotagged tweets that mention COVID-19. To protect Twitter users' privacy and to comply with Twitter's data-sharing policy, we aggregated tweets to their corresponding countries and subnational administrative divisions worldwide, with additional aggregation at the U.S. county level, for public data sharing in a dashboard [5]. This dashboard includes maps of COVID-19 mentions and rates (adjusted by population) by country, subnational administrative divisions, and county (in the United States) as well as graphs of daily COVID-19 mentions by country. The dashboard is updated daily.

The advantages of Twitter data—open source, big volume, in near real time—provide opportunities to study social dynamics and human behaviors central to the COVID-19 pandemic and the effectiveness of policy interventions across space and in near real time. In all such efforts, we carefully protected Twitter user privacy by presenting data only at the aggregate level so that individual users cannot be identified.

## 2. Materials and Methods

*2.1. Twitter Data Collection and Cleanup.* We have been continuously collecting the geotagged tweets for this study since January 15, 2020 by using the publicly accessible Twitter Streaming application programming interface (API) [11]. The geographical extent of the data collection covers the entire globe (But note that the data collection may be incomplete in countries that have limited or restricted access to Twitter). We stored the downloaded geotagged tweets as JavaScript Object Notation (JSON) encoded text files. Each record for a raw tweet contains over 150 information fields, including user identification (ID), geographic location (latitude/longitude), timestamp, message content, and language type. We derived locations from actual GPS locations in raw latitude and longitude and then geocoded and aggregated them to administrative units.

To identify and extract tweets with message contents related to COVID-19, we first compiled a set of predefined keywords based on earlier work [10] together with the keyword list suggested by Twitter, Inc [11]. Instead of using exact string matching for finding a keyword in a tweet, which might potentially exclude other related tweets due to slight typing variations, we employed a partial string-matching approach to extract the tweets related to COVID-19. We included a tweet in the final dataset if it contained a string with a word from the keyword list (Appendix S1). Consequently, this approach included a small portion of tweets that were irrelevant to COVID-19. To deal with that issue, we also developed a text classifier to assess the accuracy of the extracted tweets (i.e., whether they were COVID-19 related or not), which is detailed in the Data Records section.

Although Twitter data are openly accessible, sensitive personal information embedded in the tweets (e.g., user ID and geotagged location) should be protected. To do so, we anonymized all the extracted tweets by replacing the Twitter user ID with a randomly generated number. To minimize the risk of exposing the exact geo-location from which a tweet was sent, we removed geo-locations and then aggregated tweets to their national and subnational administrative divisions. First, we used the "country_code" of each tweet to assign the tweet to a country, or we coded it for subnational administrative divisions. Further, for tweets sent from the United States, we appended the second-level administrative division (county). We also performed a "point-in-polygon" spatial operation on each location against the cartographic boundaries of U.S. counties and states [12], and appended to each tweet the Federal Information Processing System (FIPS) code for the county and state. A statistical summary of the number of tweets in each county (and by county and state in the United States) by date is shown on our dashboard. The process of data preparation, analysis, and presentation is illustrated in Figure 1.

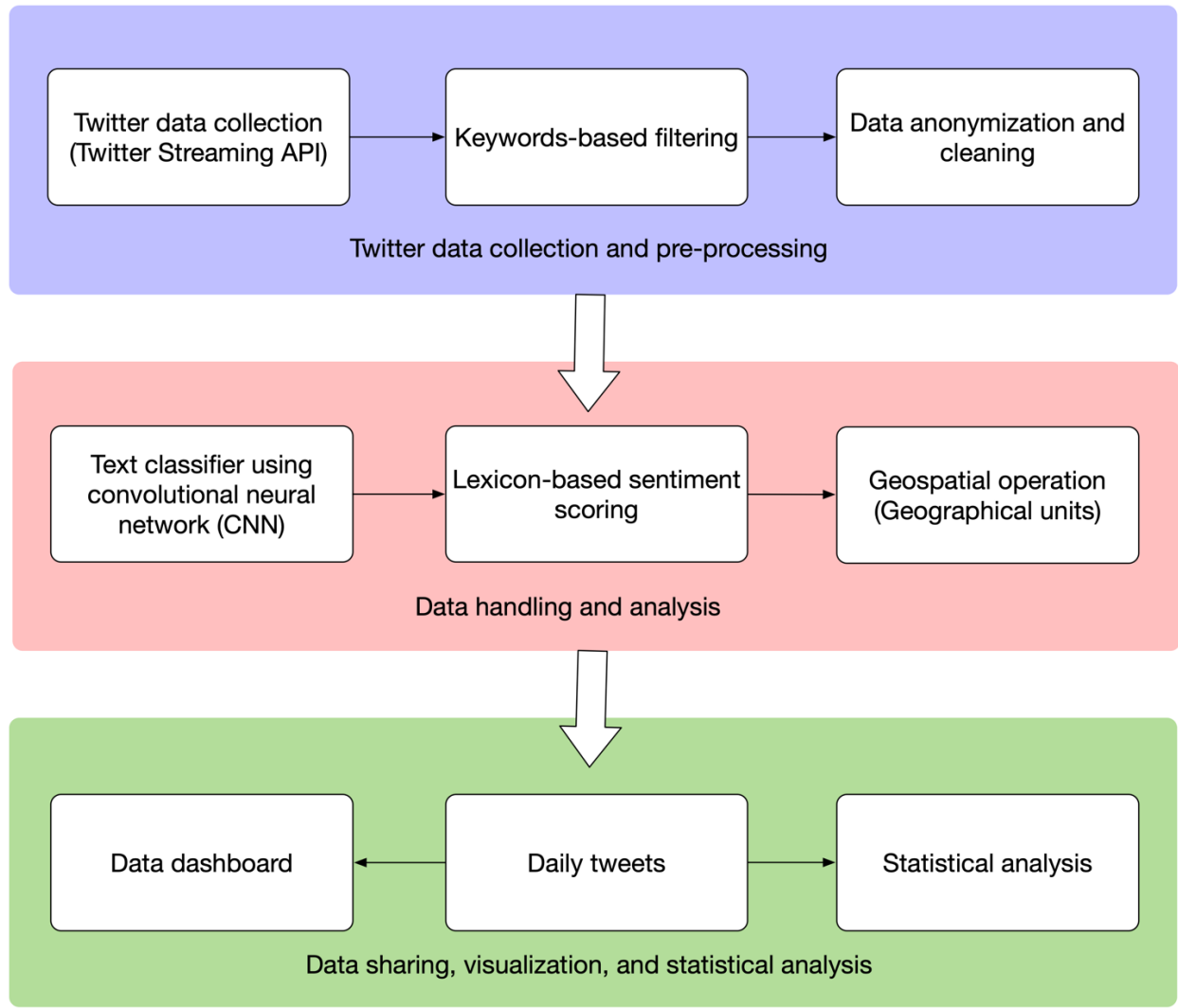

FIGURE 1. A flowchart illustrating the process of Twitter data preparation, analysis, and presentation.

*2.2. WebGIS Dashboard.* In addition to the aggregation of tweets, we developed an online Twitter dashboard using the latest WebGIS technology available for disseminating the data. Three major steps were required to build this app. The first step included background processing of the collected Twitter data in ArcGIS software and aggregation at the national and sub-country levels. We spatially joined all raw data generated by the data collection process to the country and administrative boundary files provided by ESRI. Then we calculated the tweet rate per 1000 people in each individual administrative unit. We used country and county population counts from the Socioeconomic Data and Applications Center (SEDAC) at Columbia University [13]. Then we created a visualization of each variable of interest using a white/red color pattern and quantile classification scheme. Finally, we uploaded each layer to ESRI's ArcGIS Online system as two separate web services—feature service for actual data access and tile service for faster drawing speed. Since the data feed is constantly updated, data processing is automated by the ArcGIS Python script. The resulting dataset included the rate of tweets by nation and subnational administrative divisions.

The second step involved online map design in the ArcGIS Online system. We created multiple separate web maps for this project: tweets by country as a count (Figure 2a) and a count per 1000 population (Figure 2b), tweets by subnational administrative division as a count (Figure 3a) and as a count per 1000 population (Figure 3b); and, for the United States only, tweets by subnational administrative division (state) and counties as a count and as a rate. Each map is a combination of multiple feature and tile services organized to allow the best visibility and navigation opportunities for app users. For the U.S. maps, the additional U.S. county point layer is initially hidden but is activated when a user zooms in to a particular location (Figure 4). We also created pop-up windows, defining the types of information that will be shown when a user clicks on a particular country/administrative unit on the map. We continually update web services behind all maps in the project.

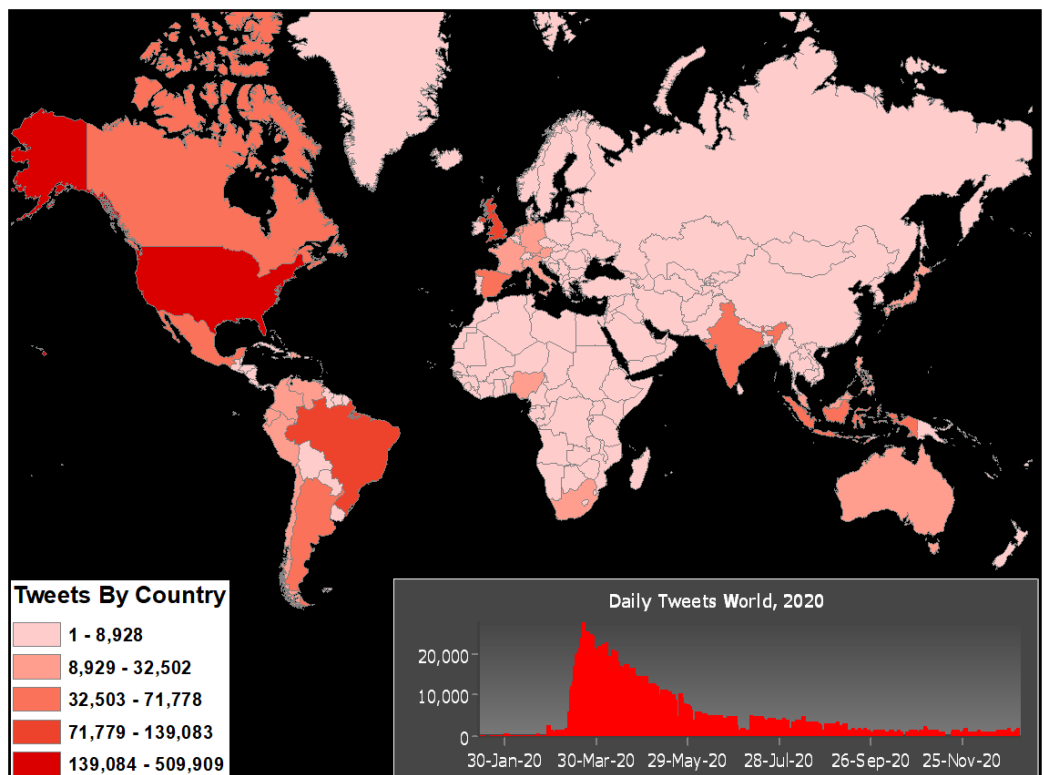

FIGURE 2a. Tweet mentions of COVID-19 by country. This is a snapshot of data as of December 31, 2020. The map shows total tweets, and the inset graph shows daily counts of tweets for the entire world since January 2020.

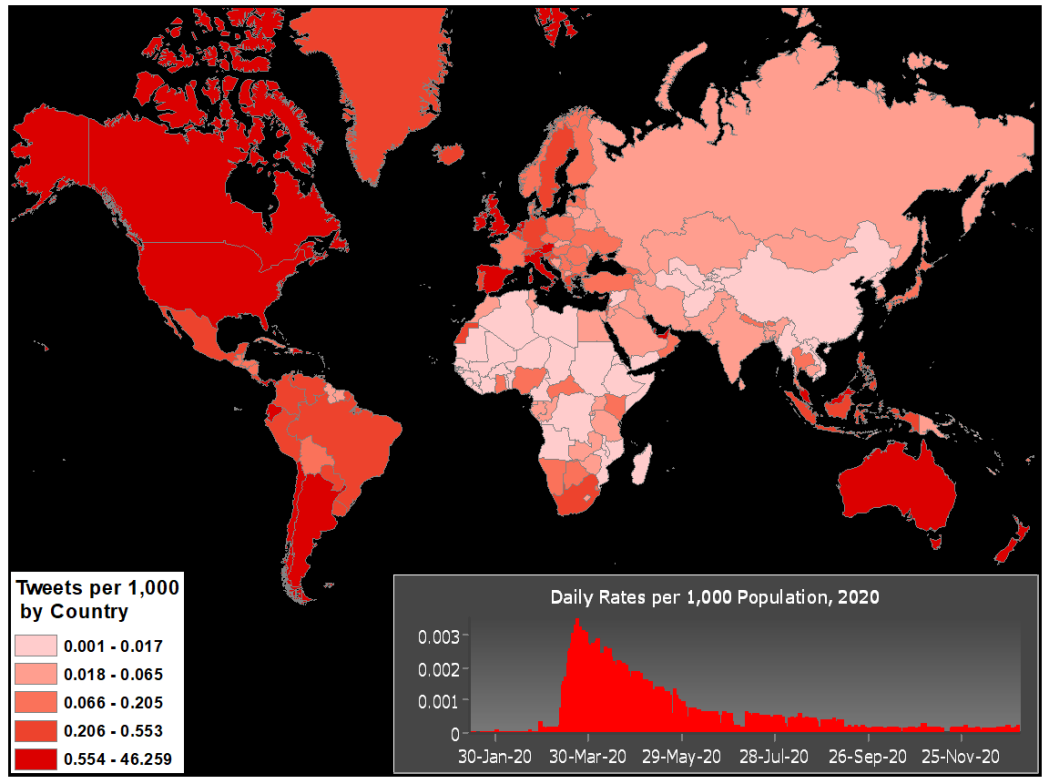

FIGURE 2b. Tweet mentions of COVID-19 (the number of tweets per 1000 people) by country. This is a snapshot of data as of December 31, 2020. The inset graph shows daily tweets for the entire world since January 2020.

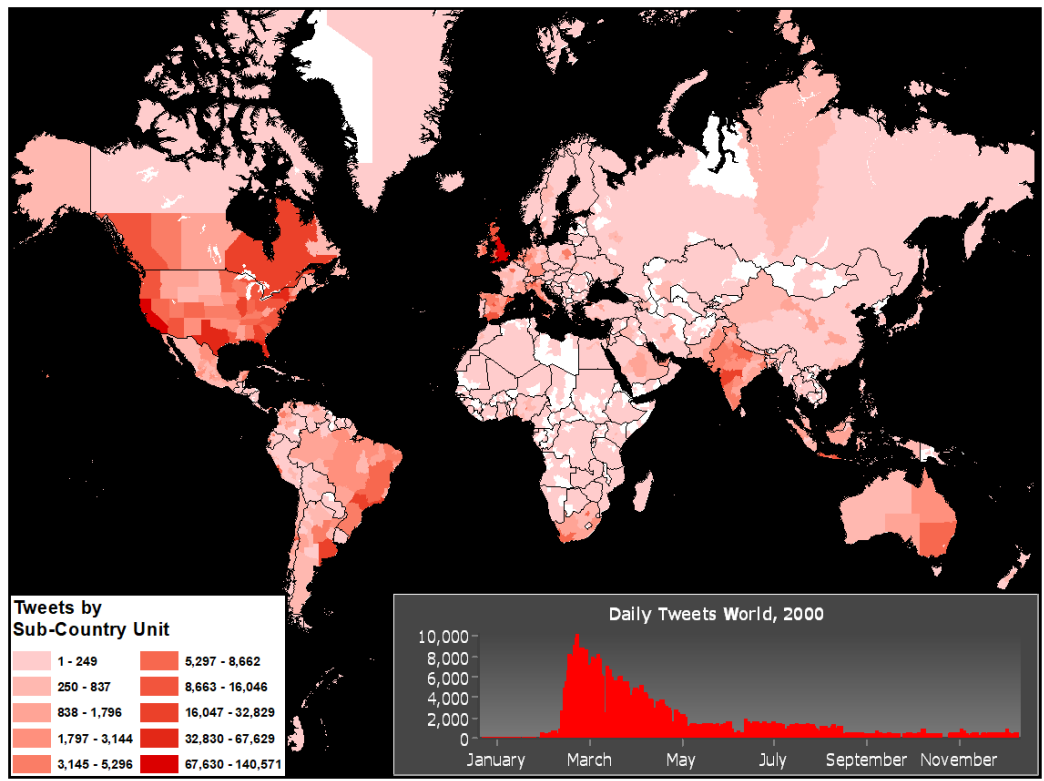

FIGURE 3a. Tweet mentions of COVID-19 by subnational administrative divisions. This is a snapshot of data as of December 31, 2020. The inset graph shows daily tweets for the entire world since January 2020.

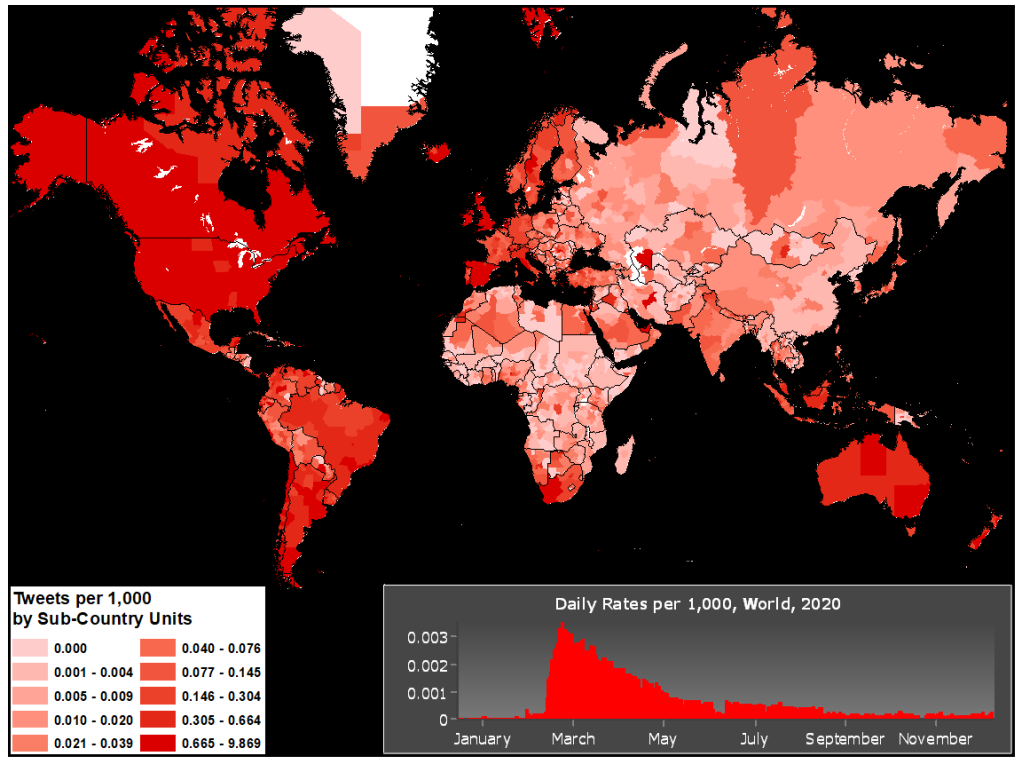

FIGURE 3b. Tweet mentions of COVID-19 (the number of tweets per 1000 people) by subnational administrative divisions. This is a snapshot of data as of December 31, 2020. The map shows total tweets per 1000 people, and the inset graph shows daily rates (the number of tweets per 1000 people) for the entire world since January 2020.

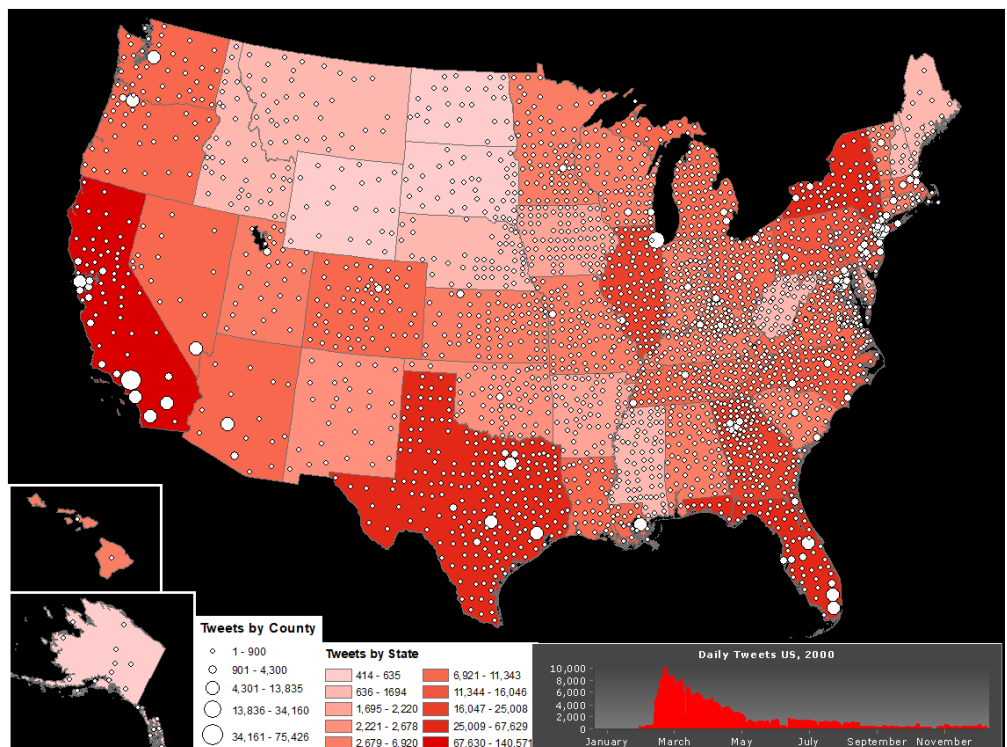

FIGURE 4. Tweet mentions of COVID-19 by states and counties in the United States. Color intensity represents the number of tweets by state, while dot size represents the number of tweets by county. This is a snapshot of data as of December 31, 2020. The map shows total tweets, and the inset graph shows daily tweets since January 2020.

In the third step, we designed a web application that visualizes the distribution of global tweet mentions of COVID-19 (http://webgis.pop.psu.edu/covid-19). We used a web template called Operations Dashboard [14], which is available for ESRI users with an ArcGIS Online Enterprise subscription. This is one of the latest additions to the ArcGIS Online system, and it allows creating complex web applications with multiple components. The app we designed for this project contains the following elements:

(1)  the maps described above,
(2)  two indicator windows with a total number of tweets and time of last update,
(3)  a list of countries with tweet counts,
(4)  a graph showing temporal distribution of tweets, and
(5)  a Notes section with the project description

In addition, the app implements a data-driven approach that connects various elements and allows automatic data adjustments. For example, selecting one of the countries on the list will automatically adjust the graph to show the data for that country.

*2.3. Data Records.* In this study, we produced two types of datasets that are openly accessible from a Harvard Dataverse repository (https://doi.org/10.7910/DVN/I2LNI5). The first dataset is a static copy of geotagged tweets related to COVID-19 stored in CSV text files collected from January 15 forward (currently through July 31, 2021). The second dataset is a CSV file summarizing the cumulative and daily numbers of tweets by country and subnational level in the same period. A live version of the data record, which will be updated on a weekly basis, can be found in the same repository. Although the data are open source, all users must agree to the terms listed in the data usage license included in the repository.

The geotagged tweets related to COVID-19 are stored as tweet-objects. Each tweet-object contains over 150 information fields, and documentation about each of the fields is provided by Twitter Inc.'s development team [15].

In addition, to comply with Twitter's sharing guidelines and to protect user privacy, we removed from the dataset identifiable information, including Twitter user IDs and coordinates of tweets. We replaced user IDs and exact coordinates by a randomly generated user ID, and tweets were located by their nation of origin, subnational level, and for the United States, the county level.

*2.4. Technical Validation.* Because we extracted the tweets related to COVID-19 based on partial-string matching, it was possible that tweets irrelevant to COVID-19 could be included. Therefore, to better assess the accuracy of the classification of tweets related to COVID-19, we applied a machine learning-based text classifier to determine whether the extracted tweets were COVID-19 related or not. For this purpose, we trained a binary text classifier utilizing the Ludwig deep learning framework [16]. To train the classifier, we prepared a training set with 1000 tweets randomly drawn from a collection of English-only datasets and manually labeled them (label_0 means irrelevant and label_1 means relevant). The training model achieved an overall accuracy of 0.88. Applying the trained classifier to the entire dataset shows that more than 87.4% (i.e., 143,338 out of 163,919) of the extracted tweets were COVID-19 related.

*2.5. Statistical Analysis.* We investigated the relationship between daily tweets and counts of COVID cases in the United States. We combined our twitter data set with publicly available counts of COVID cases for the United States from the *New York Times* (http://github.com/nytimes/covid-19-data). We created a plot with daily counts of COVID and covid keyword tweets for the United States, using both raw and a smoothed data set using a moving seven-day average of daily tweets. Based on visual examination of the plotted data, we assigned our data to three waves of COVID and then tested the Pearson's correlation between tweets and COVID case counts for each wave (January 1–June 20, June 21–September 20, September 21–December 31) at an alpha of 0.05. We used Stata (Standard edition version 17.0) to analyse our data.

## 3. Results

Figure 5 shows that COVID cases increased in a series of "waves" over time, with a spring wave from January 1–June 20, 2020 with a late April peak, a summer wave from June 21–September 20 with a late July peak, and a fall wave from September 21–December 31 that remained high at the end of the study. Tweet mentions had their highest initial spike in the spring before they declined to a plateau that corresponded to the second COVID spike before finally diminishing to near baseline levels in the fall. Both the COVID and tweet datasets showed day-to-day variation as well.

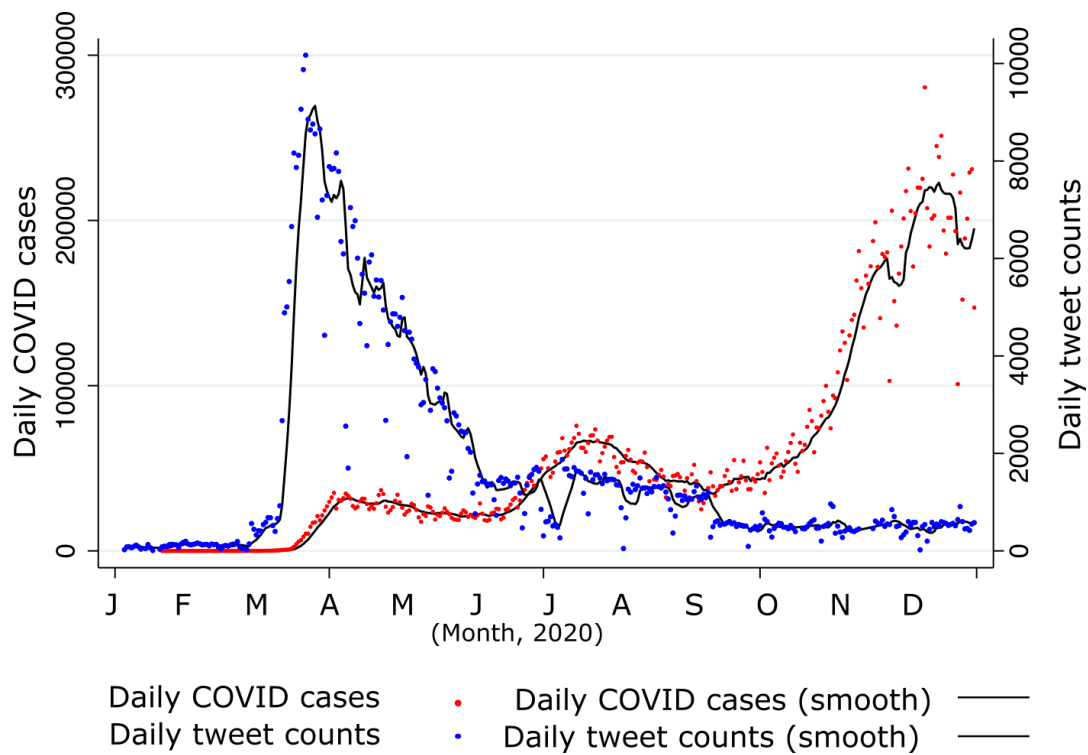

FIGURE 5: Daily counts of COVID cases and tweets in the United States, 1/1/2020 to 12/31/2020. Note: Smoothing is created with a 7-day rolling average.

There was high variability in day-to-day reporting of both COVID cases and tweet mentions, so we included a smoothed, 7-day rolling average of cases [17]. Tweet volume surged with the initial arrival of COVID in the spring of 2020, gradually decreasing to a relative plateau during the second COVID wave before dropping to small numbers in the fall. There are some missing days of Twitter data, which are indistinguishable from days with zero tweets. In addition to the decline in tweets over the course of the pandemic before they almost disappeared in the fall, there was a notable July dip—suggesting either decreased tweeting on the July 4 holiday or a possible reporting error.

We tested Pearson's correlation separately for the three waves (January 1–June 20, June 21–September 20, September 21–December 31). Pearson's correlation results are shown in Table 1. The early phase of the pandemic is evidenced by the high correlation between smoothed cases and smoothed tweets or raw cases and raw tweets. As the pandemic continued through the summer of 2020 (June 21–September 20), we continued to see a decreased but continued correlation, albeit somewhat diminished, with a correlation for the smoothed data and for the raw data. Finally, in the fall, tweet mentions overall were low, and there was no evidence of a correlation for either smoothed or raw data. Because the growth of the disease was exponential, we also explored log transformations of the COVID counts but found no meaningful difference from the correlational findings.

Table 1: Pearson correlation results for daily U.S. tweet count, rolling 7-day average, daily COVID case count, and rolling smooth COVID count for 2020 (January 1–June 20, June 21–September 20, and September 21–December 31).

|  | 1/1/2020–6/20/2020 | | 6/21/2020–9/21/2020 | | 9/21/2020–12/31/2020 | |
| --- | --- | --- | --- | --- | --- | --- |
|  | Daily tweets | Smoothed tweets | Daily tweets | Smoothed tweets | Daily tweets | Smoothed tweets |
| Daily COVID cases | 0.51*** |  | 0.33 ** |  | 0.015 |  |
| Smoothed COVID cases |  | 0.56*** |  | 0.44*** |  | –0.024 |

Note: *p <= 0.05, **p <= 0.01, ***p <= 0.001

## 4. Discussion and Conclusion

After a year and half and over 4 million deaths, the COVID-19 pandemic continues to be widespread, and its related topics continue to dominate the global media. We have created a dataset that can help researchers observe, describe, and measure patterns, and it can help suggest further research questions that could be addressed with other, similar datasets or additional, qualitative research. This data has spatial detail and is connected to administrative units for which detailed data about demographics, health, and political views are available, allowing for innumerous study questions.

We find in the United States that Twitter engagement with COVID declined over time, even as case counts increased. These quantitative measures provide an index of the signal linking tweets and COVID cases, but further work is necessary to explain the pattern of change over time. In the United States, a contentious election dominated the news cycle in fall 2020, perhaps serving as a "competing hazard" for tweet mentions. Alternatively, continued (and even increasing) case counts of disease may have created "COVID fatigue" and reduced the number of tweets. This analysis is limited to the United States, and further research in other regions is warranted.

Worldwide, COVID responses varied due to variations in vaccine availability and policy choices, and further research using this data set may provide insight into social differences in views about national and subnational level responses. In the United States, for instance, there has been notable regional variation in views about disease response including views on non-pharmaceutical interventions like masking and mandated social distancing, along with vaccine uptake rates. In future work this dataset could be combined with detailed socioeconomic and voting data to assess area level differences in engagement. Twitter data have the potential to track and measure social response to pandemic in a way that might be of use to policy makers and health officials for future planning to improve communication and support messaging strategies.

## Data and Code Availability

Datasets and code described in this manuscript are openly accessible from a Harvard Dataverse repository (https://doi.org/10.7910/DVN/I2LNI5). The first dataset is a static copy of geotagged tweets related to COVID-19 stored in CSV text files collected from January 15 forward (currently through July 31, 2021). The second dataset is a CSV file summarizing the cumulative and daily numbers of tweets by country and subnational level in the same period. A live version of the data record, which will be updated on a weekly basis, can be found in the same repository. Although the data are open source, all users must agree to the terms listed in the data usage license included in the repository.

All codes developed from this study are openly accessible from the GitHub repository named "covid-19_geo_tweets" at https://github.com/bigdata4ss/covid-19_geo_tweets. The codes include those for geotagged Twitter data collection, extraction of tweets related to COVID-19 based on keywords, spatial aggregation based on a "point-in-polygon" operation, and a statistical summary of the number of tweets by country and date. All code was developed using the programming language Python 3 [18].

## Conflicts of Interest

The authors declare that there is no conflict of interest regarding the publication of this article.

## Authors' contributions

G.C. conceived the initial concept. G.C. and J.Y. developed the research design and obtained the data. J.Y., M.L.S., and Y.B. conducted data analysis and interpretation. G.C., J.Y., M.L.S., and Y.B. contributed to the initial draft of the manuscript. All authors performed editing for the important intellectual content.

## Acknowledgments

We thank Susan H. McHale for her suggestions and comments on early versions of the work. This research was supported in part by the National Science Foundation (Awards # SES-1823633 and # OPP-2032790), the USDA National Institute of Food and Agriculture and Multistate Research Project #PEN04623 (Accession #1013257), the Eunice Kennedy Shriver National Institute of Child Health and Human Development (Award # P2C HD041025), and the Social Science Research Institute, Population Research Institute, and Institute for Computational and Data Sciences of the Pennsylvania State University.

## Supplementary Materials

*Supplementary 1.* Supporting Information S1: this file contains the list of Twitter keywords based on which we extracted the COVID related tweets.

# Supplementary Materials

Supporting Information S1: List of Twitter Keywords

#Coronavirusmexico
#covid2019
#coronavirususa
#covid_19uk
#covid-19uk
#Briefing_COVID19
#coronaapocolypse
#coronavirusbrazil
#marchapelocorona
#coronavirusbrasil
#coronaday
#coronafest
#coronavirusu
#covid2019pt
#COVID19PT
#caronavirususa
#covid19india
#caronavirusindia
#caronavirusoutbreak
#caronavirus
carona virus
#2019nCoV
2019nCoV
#codvid_19
#codvid19
#conronaviruspandemic
#corona
corona
corona vairus
corona virus
#coronadeutschland
#Coronaferien
#coronaflu
#coronaoutbreak
#coronapandemic
#Coronapanik
#coronapocalypse
#CoronaSchlager
#coronavid19
#coronavid19
#Coronavirus
Coronavirus

#coronavirusargentina
#coronavirusbrasil
#CoronaVirusCanada
#coronaviruschile
#coronaviruscolombia
#CoronaVirusDE
#coronavirusecuador
#CoronavirusEnColombia
#coronavirusespana
CoronavirusFR
#CoronavirusFR
#coronavirusIndonesia
#Coronavirusireland
#CoronaVirusIreland
#coronavirusmadrid
#coronavirusmexico
#coronavirusnobrasil
#coronavirusnyc
#coronavirusoutbreak
#coronavirusoutbreak
#coronaviruspandemic
#coronavirusperu
#coronaviruspuertorico
#coronavirusrd
#coronavirustruth
#coronavirusuk
coronavirusupdate
#coronavirusupdates
#coronavirusuruguay
coronga virus
corongavirus
#Corvid19virus
#covd19
#covid
covid
#covid
covid
covid 19
#covid_19
#covid_19
Covid_19
#COVID_19uk
#covid19
Covid19
Covid19_DE
#covid19Canada

Covid19DE
Covid19Deutschland
#covid19espana
#covid19france
#covid19Indonesia
#covid19ireland
#covid19uk
#covid19usa
#covid2019
#ForcaCoronaVirus
#infocoronavirus
#kamitidaktakutviruscorona
#nCoV
nCoV
#ncov2019
nCoV2019
NeuerCoronavirus
#NeuerCoronavirus
Nouveau coronavirus
#NouveauCoronavirus
novel coronavirus
#NovelCorona
novelcoronavirus
#novelcoronavirus
#NovelCoronavirus
#NuovoCoronavirus
#ohiocoronavirus
#PánicoPorCoranovirus
#SARSCoV2
#SARSCoV2
the coronas
#thecoronas
#trumpdemic
Virus Corona
#viruscorona
#فيروس_كورونا
#كورونا
#كورونا_الجديد
#कोरोना
कोरोना
कोरोना वायरस
#कोरोना_वायरस
코로나
#코로나
#코로나19

코로나바이러스
#코로나바이러스
コロナ
#コロナ
#コロナウイルス
加油武汉
#加油武汉
#新冠病毒
新冠病毒
#新冠肺炎
新冠肺炎
#新型コロナウイルス
#新型冠状病毒
新型冠状病毒
武汉加油
#武汉加油
#武汉疫情
#武汉肺炎
武汉肺炎
#武漢肺炎
武漢肺炎
疫情
#疫情
#CoronaAlert
#coronavirusUP
#coronavirustelangana
#coronaviruskerala
#coronavirusmumbai
#coronavirusdelhi
#coronavirusmaharashtra
#coronavirusinindia
वूहान
#covid_19ind
#covid19india
coronavirus india
#coronavirusindia
#कोविड_19
#कोविड-१९
#कोरोनावायरस
#bayarealockdown

#stayathomechallenge
#stayhomechallenge
#quarantinelife
#dontbeaspreader
#stayhomechallenge
#howtokeeppeoplehome
#togetherathome
alcool em gel
alcool gel
#alcoolemgel
#alcoolgel
#avoidcrowds
bares cerrados
bares fechados
bars closed
#canceleverything
#CerradMadridYa
clases anuladas
#CLOSENYCPUBLICSCHOOLS
#confinementtotal
#CONVID19
#CoronavirusESP
cuarentena
#cuarentena
#CuarentenaCoronavirus
#cuarentenaYA
dont touch ur face
dont touch your face
#DontBeASpreader
#donttouchyourface
escolas fechadas
escolas fechando
escolas sem aula
escolas sem aulas
#euficoemcasa
evitar el contagio
#ficaemcasa
flatten the curve
flattening the curve
#flatteningthecurve
#flattenthecurve
#FrenaLaCurva
Hand sanitizer
#Handsanitizer
#HoldTheVirus
lava tu manos

#lavatumanos
lave as maos
#laveasmaos
#lockdown
lockdown
#pandemic
pandemic
#panicbuying
#panickbuing
quarantaine
#quarantine
quarantine
#QuarantineAndChill
quarantined
quarentena
#quarentena
#quarentine
#quarentined
quarentined
#quarentinelife
#quedateencasa
#remotework
#remoteworking
restaurantes cerrados
restaurantes fechados
restaurants closed
#selfisolating
#SiMeContagioYo
social distancing
#socialdistance
#socialdistancing
#socialdistancingnow
#socialdistnacing
#stayathome
#stayathome
#stayhome
#stayhome
#stayhomechallenge
#stayhomesavelives
#StayTheFHome
#StayTheFuckHome
#suspendanlasclases
teletrabajo
#teletrabajo
#ToiletPaperApocalypse
#toiletpaperpanic

trabajadores a la calle
trabajar desde casa
#trabajardesdecasa
trabalhando de casa
trabalhar de casa
wash ur hands
wash your hands
#washurhands
#washyourhands
#WashYourHandsAgain
#wfh
work from home
#workfromhome
working from home
#workingfromhome
#yomequedoencasa
#covid19
#coronavirus
#stayhome
#socialdistancingworks
#socialdistancing
#socialdistancing2020
#socialdistance
#practicesocialdistancing
#socialdistancingsaveslives
#keepsocialdistance
#maintainsocialdistance
#socialdistancingfail
#maintainsocialdistancing
#covid
#covid_19
#stayathome
#lockdown
#staysafe
#corona
#quarantine
#stayhomesavelives
#pandemic
#flattenthecurve
#coronaviruspandemic
#washyourhands
#selfisolation
#socialdistance
#coronavirusoutbreak
#stayhomestaysafe
#socialdistancingnow

#quarantinelife
#physicaldistancing
#coronalockdown
#socialdistanacing
#coronaviruslockdown
#stopthespread
#coronavirususa
#stayhealthy
#staysafestayhome
#social_distancing
#selfquarantine
#stayathomeandstaysafe
covid19
coronavirus
stayhome
socialdistancingworks
socialdistancing
socialdistancing2020
socialdistance
practicesocialdistancing
socialdistancingsaveslives
keepsocialdistance
maintainsocialdistance
socialdistancingfail
maintainsocialdistancing
covid
covid_19
stayathome
lockdown
staysafe
quarantine
stayhomesavelives
pandemic
flattenthecurve
coronaviruspandemic
washyourhands
selfisolation
socialdistance
coronavirusoutbreak
stayhomestaysafe
socialdistancingnow
quarantinelife
physicaldistancing
coronalockdown
socialdistanacing
coronaviruslockdown

stopthespread
coronavirususa
stayhealthy
staysafestayhome
social_distancing
selfquarantine
stayathomeandstaysafe